# Nightside condensation of iron in an ultra-hot giant exoplanet

David Ehrenreich[1], Christophe Lovis[1], Romain Allart[1], María Rosa Zapatero Osorio[2], Francesco Pepe[1], Stefano Cristiani[3], Rafael Rebolo[4], Nuno C. Santos[5,6], Francesco Borsa[7], Olivier Demangeon[5], Xavier Dumusque[1], Jonay I. González Hernández[4], Núria Casasayas-Barris[4], Damien Ségransan[1], Sérgio Sousa[5], Manuel Abreu[8,9], Vardan Adibekyan[5], Michael Affolter[10], Carlos Allende Prieto[4], Yann Alibert[10], Matteo Aliverti[7], David Alves[8,9], Manuel Amate[4], Gerardo Avila[11], Veronica Baldini[3], Timothy Bandy[10], Willy Benz[10], Andrea Bianco[7], Émeline Bolmont[1], François Bouchy[1], Vincent Bourrier[1], Christopher Broeg[10], Alexandre Cabral[8,9], Giorgio Calderone[3], Enric Pallé[4], H. M. Cegla[1], Roberto Cirami[3], João M. P. Coelho[8,9], Paolo Conconi[7], Igor Coretti[3], Claudio Cumani[11], Guido Cupani[3], Hans Dekker[11], Bernard Delabre[11], Sebastian Deiries[11], Valentina D'Odorico[3,12], Paolo Di Marcantonio[3], Pedro Figueira[13,5], Ana Fragoso[4], Ludovic Genolet[1], Matteo Genoni[7], Ricardo Génova Santos[4], Nathan Hara[1], Ian Hughes[1], Olaf Iwert[11], Florian Kerber[11], Jens Knudstrup[11], Marco Landoni[7], Baptiste Lavie[1], Jean-Louis Lizon[11], Monika Lendl[1,14], Gaspare Lo Curto[13], Charles Maire[1], Antonio Manescau[11], C. J. A. P. Martins[5,15], Denis Mégevand[1], Andrea Mehner[13], Giusi Micela[16], Andrea Modigliani[11], Paolo Molaro[3,17], Manuel Monteiro[5], Mario Monteiro[5,6], Manuele Moschetti[7], Eric Müller[11], Nelson Nunes[8], Luca Oggioni[7], António Oliveira[8,9], Giorgio Pariani[7], Luca Pasquini[11], Ennio Poretti[7,18], José Luis Rasilla[4], Edoardo Redaelli[7], Marco Riva[7], Samuel Santana Tschudi[13], Paolo Santin[3], Pedro Santos[8,9], Alex Segovia Milla[1], Julia V. Seidel[1], Danuta Sosnowska[1], Alessandro Sozzetti[19], Paolo Spanò[7], Alejandro Suárez Mascareño[4], Hugo Tabernero[2,5], Fabio Tenegi[4], Stéphane Udry[1], Alessio Zanutta[7], Filippo Zerbi[7]

**Ultra-hot giant exoplanets receive thousands of times Earth's insolation[1,2]. Their high-temperature atmospheres (>2,000 K) are ideal laboratories for studying extreme planetary climates and chemistry[3-5]. Daysides are predicted to be cloud-free, dominated by atomic species[6] and substantially hotter than nightsides[5,7,8]. Atoms are expected to recombine into molecules over the nightside[9], resulting in different day-night chemistry. While metallic elements and a large temperature contrast have been observed[10-14], no chemical gradient has been measured across the surface of such an exoplanet. Different atmospheric chemistry between the day-to-night ("evening") and night-to-day**

[1]Observatoire astronomique de l'Université de Genève, 51 chemin des Maillettes, 1290 Versoix, Switzerland. [2]Centro de Astrobiología (CSIC-INTA), Carretera de Ajalvir km 4, E-28850 Torrejón de Ardoz, Madrid, Spain. [3]INAF Osservatorio Astronomico di Trieste, via G. Tiepolo 11, 34143 Trieste, Italy. [4]Instituto de Astrofísica de Canarias, Vía Láctea s/n, 38205 La Laguna, Tenerife, Spain. [5]Instituto de Astrofísica e Ciências do Espaço, Universidade do Porto, CAUP, Rua das Estrelas, 4150-762 Porto, Portugal. [6]Departamento de Física e Astronomia, Faculdade de Ciências, Universidade do Porto, Rua do Campo Alegre, 4169-007 Porto, Portugal. [7]INAF Osservatorio Astronomico di Brera, Via E. Bianchi 46, 23807 Merate, Italy. [8]Instituto de Astrofísica e Ciências do Espaço, Universidade de Lisboa, Edifício C8, 1749-016 Lisboa, Portugal. [9]Departamento de Física da Faculdade de Ciências da Univeridade de Lisboa, Edifício C8, 1749-016 Lisboa, Portugal. [10]Physikalisches Institut & Center for Space and Habitability, Universität Bern, Gesellschaftsstrasse 6, 3012 Bern, Switzerland. [11]European Southern Observatory, Karl-Schwarzschild-Str. 2, 85748 Garching bei München, Germany. [12]Scuola Normale Superiore, P.zza dei Cavalieri, 7, 56126 Pisa, Italy. [13]European Southern Observatory, Alonso de Córdova 3107, Vitacura, Casilla 19001, Santiago de Chile, Chile. [14]Space Research Institute, Austrian Academy of Sciences, Schmiedlstr. 6, A-8042 Graz, Austria. [15]Centro de Astrofísica da Universidade do Porto, Rua das Estrelas, 4150-762 Porto, Portugal. [16]INAF Osservatorio Astronomico di Palermo, Piazza del Parlamento 1, 90134 Palermo, Italy. [17]Institute for Fundamental Physics of the Universe, Via Beirut 2, 34151 Miramare, Trieste, Italy. [18]Fundación Galileo Galilei, INAF, Rambla José Ana Fernandez Pérez 7, 38712 Breña Baja, Spain. [19]INAF Osservatorio Astrofisico di Torino, Via Osservatorio 20, 10025 Pino Torinese, Italy.

("morning") terminators could, however, be revealed as an asymmetric absorption signature during transit[4,7,15]. Here, we report the detection of an asymmetric atmospheric signature in the ultra-hot exoplanet WASP-76b. We spectrally and temporally resolve this signature thanks to the combination of high-dispersion spectroscopy with a large photon-collecting area. The absorption signal, attributed to neutral iron, is blueshifted by $-11\pm0.7$ km s$^{-1}$ on the trailing limb, which can be explained by a combination of planetary rotation and wind blowing from the hot dayside[16]. In contrast, no signal arises from the nightside close to the morning terminator, showing that atomic iron is not absorbing starlight there. Iron must thus condense during its journey across the nightside.

[1]Observatoire astronomique de l'Université de Genève, 51 chemin des Maillettes, 1290 Versoix, Switzerland. [2]Centro de Astrobiología (CSIC-INTA), Carretera de Ajalvir km 4, E-28850 Torrejón de Ardoz, Madrid, Spain. [3]INAF Osservatorio Astronomico di Trieste, via G. Tiepolo 11, 34143 Trieste, Italy. [4]Instituto de Astrofísica de Canarias, Vía Láctea s/n, 38205 La Laguna, Tenerife, Spain. [5]Instituto de Astrofísica e Ciências do Espaço, Universidade do Porto, CAUP, Rua das Estrelas, 4150-762 Porto, Portugal. [6]Departamento de Física e Astronomia, Faculdade de Ciências, Universidade do Porto, Rua do Campo Alegre, 4169-007 Porto, Portugal. [7]INAF Osservatorio Astronomico di Brera, Via E. Bianchi 46, 23807 Merate, Italy. [8]Instituto de Astrofísica e Ciências do Espaço, Universidade de Lisboa, Edifício C8, 1749-016 Lisboa, Portugal. [9]Departamento de Física da Faculdade de Ciências da Univeridade de Lisboa, Edifício C8, 1749-016 Lisboa, Portugal. [10]Physikalisches Institut & Center for Space and Habitability, Universität Bern, Gesellschaftsstrasse 6, 3012 Bern, Switzerland. [11]European Southern Observatory, Karl-Schwarzschild-Str. 2, 85748 Garching bei München, Germany. [12]Scuola Normale Superiore, P.zza dei Cavalieri, 7, 56126 Pisa, Italy. [13]European Southern Observatory, Alonso de Córdova 3107, Vitacura, Casilla 19001, Santiago de Chile, Chile. [14]Space Research Institute, Austrian Academy of Sciences, Schmiedlstr. 6, A-8042 Graz, Austria. [15]Centro de Astrofísica da Universidade do Porto, Rua das Estrelas, 4150-762 Porto, Portugal. [16]INAF Osservatorio Astronomico di Palermo, Piazza del Parlamento 1, 90134 Palermo, Italy. [17]Institute for Fundamental Physics of the Universe, Via Beirut 2, 34151 Miramare, Trieste, Italy. [18]Fundación Galileo Galilei, INAF, Rambla José Ana Fernandez Pérez 7, 38712 Breña Baja, Spain. [19]INAF Osservatorio Astrofisico di Torino, Via Osservatorio 20, 10025 Pino Torinese, Italy.

Two transits of the short-period (1.81 day) giant exoplanet WASP-76b[17-19] were observed on 2 September 2018 (epoch 1) and 30 October 2018 (epoch 2) with the Echelle Spectrograph for Rocky Exoplanets and Stable Spectroscopic Observations (ESPRESSO) at the European Southern Observatory Very Large Telescope (VLT) located on Cerro Paranal, Chile (see the observation log in Methods and Extended Data Fig. 1). ESPRESSO is a fibre-fed, stabilised and high-resolution spectrograph[20] able to collect the light from any combination of the four VLT 8-metre Unit Telescopes (UTs). During each epoch, we acquired data with UT3 only. We used the single high-resolution 2×1-binning mode ($\langle \lambda/\Delta\lambda \rangle$ = 138,000), exposure times of 600 s and 300 s with the slow read-out mode to record 35 and 70 spectra of the bright (V=9.5), F7 star WASP-76 during epochs 1 and 2, respectively. The ESPRESSO pipeline version 1.3.2 was used to produce 1D spectra and cross-correlation functions (CCFs). The CCFs, which are average stellar line profiles, were extracted using an F9 mask over a velocity range of [−150,+150] km s$^{-1}$, with a step of 0.5 km s$^{-1}$. This stellar mask contains 4,653 spectral lines in the wavelength range between 380 and 788 nm covered by ESPRESSO; most of the spectroscopic information in the mask is contained in electronic transitions of neutral iron (Fe; see Methods).

We used the spectra to revise the stellar parameters and the CCF peak position to monitor the radial velocities of the integrated stellar disc during the transit of the planet (see Methods and Extended Data Table 1). The planet blocks different parts of the rotating stellar surface during the transit, resulting in a spectroscopic anomaly known for eclipsing binaries and exoplanets as the Rossiter-McLaughlin effect (e.g. ref. [21]). The shape of the anomaly observed for WASP-76b (Fig. 1a) shows that the planet orbit is prograde, and its orbital spin is approximately aligned with the rotational spin of the star. For this bright star, ESPRESSO yields an average photon-noise–limited precision of 70 cm s$^{-1}$ and 85 cm s$^{-1}$ per 10 min and 5 min exposure in epochs 1 and 2, respectively. This precision is high enough to reveal a small "bump" towards positive radial velocities occurring 30 min after mid-transit during each epoch. We could also find it *a posteriori* in previous data taken with the HARPS spectrograph (Fig. 1a).

We retrieved the CCFs of the stellar surface occulted by the planet during the transit[22] (the "local" CCFs). These local CCFs, shifted to the stellar rest frame, are displayed for individual epochs in Extended Data Fig. 4. They exhibit a dark slanted feature called the Doppler shadow[1]. The radial velocity of the Doppler shadow across the transit is related to the stellar projected rotational velocity $v_{eq} \sin i_\star$ (where $i_\star$ is the stellar inclination) and the projected spin-orbit obliquity $\lambda$[22]. For each in-transit exposure, we fitted the Doppler shadow with a Voigt profile. The radial velocity of the peak corresponds to the local velocity of the stellar surface behind the planet (see Methods and Fig. 1b). A fit to the data with a stellar surface model described in Methods yields $\lambda = 61\pm7°$ and a slow projected stellar rotation ($v_{eq} \sin i_\star = 1.5 \pm 0.3$ km s$^{-1}$). The system geometry is sketched in Fig. 1c.

We removed the Doppler shadow by subtracting the Voigtian fits to the data for each exposure and searched for faint planetary signals in the residuals. Any signal tied to the planet should move with a velocity close to the planet Keplerian velocity, which varies between −53 and +53 km s$^{-1}$ during transit. The residual maps, shown for both epochs combined and each epoch separately in Fig. 2a and Extended Data Fig. 5, respectively, show a slanted residual *absorption* signature close to the expected radial velocity of the planet. In contrast with the negative Doppler shadow, the slanted signature appears as a positive signal. The gap between −0.2 h and +0.7 h around mid-transit results from the subtraction of the Doppler shadow.

The absorption signal arises from the cross-correlation of the planetary atmosphere transmission spectrum with the stellar mask dominated by atomic iron lines. Therefore, it traces the presence of atomic iron in the atmosphere of WASP-76b, as found in the atmospheres of other ultra-hot gas giants[10-12,23]. The planetary absorption overlaps the Doppler shadow at the same transit phase as the "bump" in Fig. 1a, meaning that this anomaly in the Rossiter-McLaughlin effect is due to the presence of a hot planetary atmosphere, as described in previous works[23,24]. We excluded this phase range from our analysis.

We studied the absorption signature in the planetary rest frame (Fig. 2b), using our newly derived orbital solution (see Methods). It appears asymmetric and mostly blueshifted. We fitted the absorption signal with Gaussians to retrieve its amplitude, radial velocity and full-width-at-half-maximum (FWHM) as a function of time (see Extended Data Fig. 7).

The radial velocity of the planetary signature (Extended Data Fig. 7a) is slightly blueshifted (between 0 and −5 km s$^{-1}$) at ingress. It progressively blueshifts down to about −11±0.7 km s$^{-1}$, reached at −0.4 h from mid-transit. It remains blueshifted until the end of transit. The amplitude can be converted to a differential transit depth $\delta_{atm}$ caused by the atmospheric absorption (Extended Data Fig. 7c). We measure a mean absorption signal of 494±27 ppm (weighted by the uncertainties), which corresponds to the absorption by ~1.8 atmospheric scale heights calculated assuming a dayside temperature of 2,693 K[25]. During the first half of the transit, the signal contrast is 434±32 ppm close to the planet rest velocity; it increases (with a significance of 3.5σ) after +1 h from mid-transit, up to 628±49 ppm towards the end of the transit, where the signal is significantly blueshifted. An atmospheric signal centred at non-zero radial velocities indicates a motion of the absorber in the planet rest frame, typically due to winds[16]. The blueshifted signal is highly significant compared to the absence of a signal (0±49 ppm) observed at the same time around 0 km s$^{-1}$ at the 9-σ confidence level (628 ppm / (49 ppm √2) ≈ 9σ). Finally, the FWHM (Extended Data Fig. 6b) has a weighted-mean value of 8.6±0.7 km s$^{-1}$. This width could result from the combination of the tidally-locked planetary rotation (5.3 km s$^{-1}$), thermal broadening (~0.7 km s$^{-1}$) and turbulent motions due to winds.

Three-dimensional global climate models will be needed to fully take advantage of these spectrally and temporally resolved figures. Meanwhile, we can craft a toy-model to qualitatively understand the temporal evolution of the atmospheric signature. Ultra-hot gas giants have day-side temperatures commensurable with the surface of cool stars. There, most molecules should thermally dissociate, resulting in a composition dominated by atoms and ions. These partially ionised atmospheres could give rise to frictional drag caused by Lorentz forces, slowing down the characteristic time scale for heat advection[7,8]. Consequently, heat redistribution to the nightside should be relatively inefficient and the day-night temperature contrast correspondingly large. This could also result in day-to-night, longitudinally-symmetric winds[14,16] and recombination and dissociation of atoms and molecules at the evening and morning terminators[9], respectively. Our model is sketched in Fig. 3 and makes use of the following ingredients: (i) A tidally-locked rotation red- and blue-shifting the signal at the leading and trailing limb, respectively, by ±5.3 km s$^{-1}$; (ii) an absorber (neutral iron) found on the hottest part of the planet dayside (the "hot spot") and absent from the colder nightside; (iii) a longitudinal offset of the hot spot towards the evening terminator (this is in tension with the existence of strong drag forces but is necessary to explain the asymmetry between the beginning and end of transit); (iv) a uniform day-to-night wind, previously observed for HD 209458b[16]. We assumed the day-to-night wind imprints a −5.3 km s$^{-1}$ blueshift at both limbs, therefore compensating the planetary rotation redshift at the leading limb (shifting the signal towards 0 km s$^{-1}$) while increasing the blueshift at the trailing limb to −10.6 km s$^{-1}$. This wind speed lies at the upper bound of the range expected by theory[7,15,26,27]; however, most existing studies have focused on planets cooler than WASP-76b. Finally, we considered (v) the variation of the angle $\zeta = 2 \arcsin R_\star/a$ between the planet terminator and the line of sight, where $a$ is the semi-major axis of the planet. Due to the tight orbit of WASP-76b, this angle varies by 29.4° during transit. This effect brings the region containing atomic iron in and out of view during the transit.

The transit of our model in front of the star follows the three-step scenario depicted in Fig. 3: (i) At ingress, only the leading limb contributes to the signal; there, the line-of-sight crosses a fraction of the dayside containing iron atoms that absorb the starlight between 0 and −5 km s$^{-1}$. The trailing limb, entering the stellar disc, also contains iron atoms that starts to blueshift the signal. (ii) As soon as ζ is large enough to take the patch of absorbing iron atoms out of view from the leading limb, its contribution disappears. Only the trailing limb now contributes to the signal; the planetary rotation plus day-to-night winds brings it to −10.6 km s$^{-1}$. Planetary rotation could still broaden the signal after the disappearance of the leading limb signal because an absorption crescent on the trailing limb still features differential rotation from the pole to the equator. The signal then keeps a constant blueshift until egress (iii) while its absorption depth increases as the hot spot comes into view in the trailing limb. We can quantify the temperature increase between the evening terminator and the hot spot by considering that $\delta_{atm}^{day}/\delta_{atm}^{eve} = T^{day}/T^{eve}$ (see Methods). Given the values quoted above, the temperature must increase by a factor of 1.5±0.2 across the evening terminator towards the dayside. This is consistent with expectations for a strong day-night temperature contrast for ultra-hot gas giants[28] and yields an evening temperature of 1,795±242 K. Meanwhile, the egress of the leading limb does not have any apparent impact, strengthening its lack of contribution.

We conclude from this that neutral iron atoms must be present on the dayside and evening terminator, but much less abundant or even absent from the nightside and morning terminator. Therefore, iron must condense across the nightside. Nightside clouds have been suggested from thermal phase curves of hot gas giants[28,29]. On WASP-76b and similarly hot planets, these clouds could be made out of iron droplets, since liquid iron is the most stable high-temperature iron-bearing condensate[30]. Hence, it could literally rain iron on the nightside of WASP-76b.

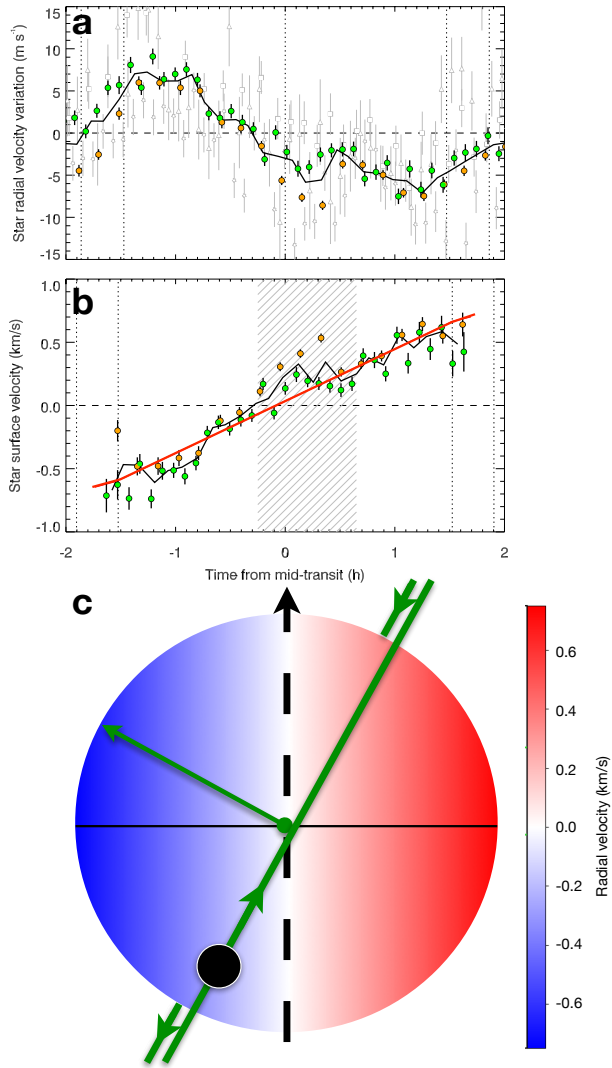

**Figure 1 | Rossiter-McLaughlin effect of WASP-76b. a,** "Classical" analysis of the effect showing the radial velocities integrated over the whole stellar disc for the ESPRESSO epoch 1 (orange), epoch 2 (green), both epochs combined (black thick curve) and 3 previous transits observed with HARPS (grey symbols; ref. [19]). **b**, "Reloaded" analysis of the effect showing the stellar surface velocities behind the disc of the planet. The red curve is a fit with a stellar surface model assuming solid-body rotation. Vertical dotted lines indicate the transit contacts and mid time. The hatched area delimits the times when the planet absorption signal crosses the Doppler shadow. The $1\sigma$ uncertainties have been propagated accordingly from the errors calculated by the ESPRESSO pipeline. Velocity scales are in the stellar rest frame. **c**, Sketch of the WASP-76 system (to scale) as seen from Earth. Arrows show the projected spin axes of the planetary orbit (green) and the star (black).

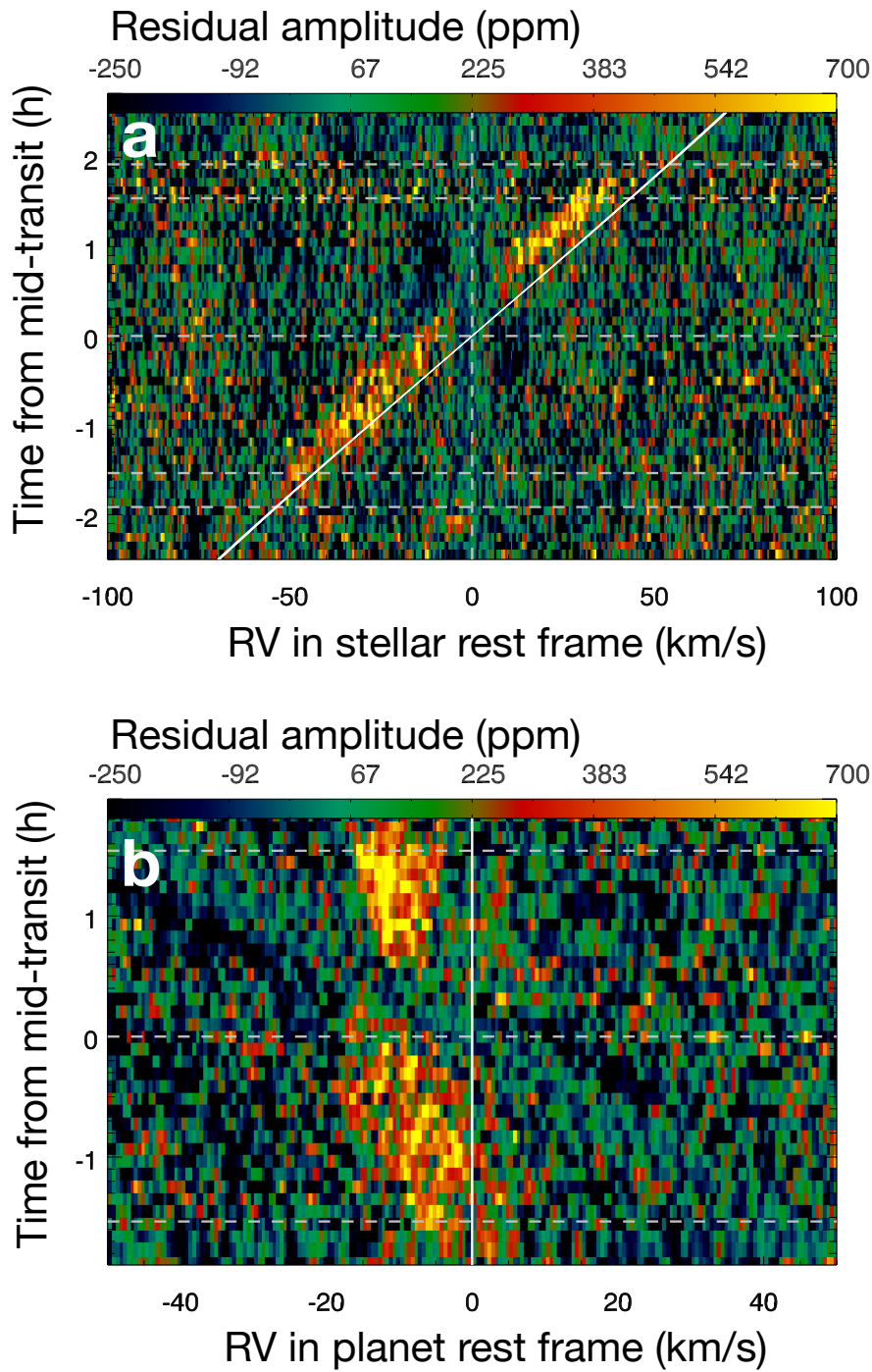

**Figure 2 | Planet absorption signature. a,** In the stellar rest frame, the planetary absorption signal appears close to the expected Keplerian of the planet, superimposed in white with its 1σ uncertainty. Transit contacts are shown by white horizontal dashed lines. The gap around 0 km s$^{-1}$ corresponds to the position of the Doppler shadow before its subtraction. **b,** In the planet rest frame, the shimmer is asymmetric and progressively blueshifts after ingress.

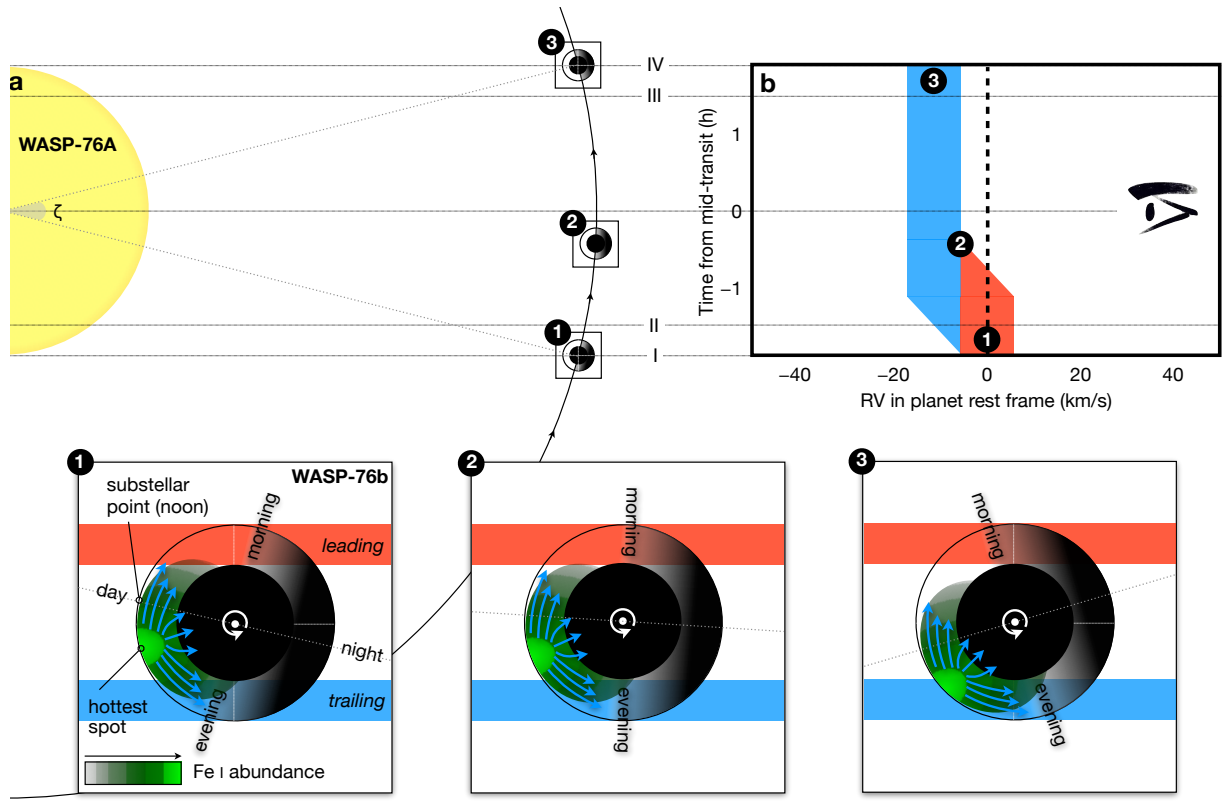

**Figure 3 | Polar view of the WASP-76 system. a,** The star WASP-76 and planet WASP-76b are represented to scale in size and distance. The planet is shown at different transit stages, with the transit contacts I, II, III and IV. During transit, the angle ζ between the planet terminator and the line of sight (dashed line in the middle) changes by $2 \arcsin R_\star/a = 29.4°$, where $a$ is the semi-major axis. **b,** Sketch of the absorption signature observed during transit, in the planet rest frame. The numbers refer to the insets. **(1)** During ingress, iron on the dayside is visible through the leading limb and creates an absorption around 0 km s$^{-1}$. The trailing limb enters the stellar disc and progressively blueshifts the signal. **(2)** The signal around 0 km s$^{-1}$ disappears as soon as no more iron is visible in the leading limb. Only the trailing limb contributes to the signal, which remain blueshifted around −11 km s$^{-1}$. **(3)** The signal remains at this blueshifted velocity until the end of the transit.

**ACKNOWLEDGEMENTS**

We thank G. Fu for sharing information regarding the binary companion of WASP-76A and D. Kitzmann for discussion about how iron can condense. This project has received funding from the European Research Council (ERC) under the European Union's Horizon 2020 research and innovation programme (project FOUR ACES; grant agreement No 724427). It has also been carried out in the frame of the National Centre for Competence in Research PlanetS supported by the Swiss National Science Foundation (SNSF). This work was supported by FCT/MCTES through national funds and by FEDER—Fundo Europeu de Desenvolvimento Regional through COMPETE2020—Programa Operacional Competitividade e Internacionalização by these grants: UID/FIS/04434/2019; PTDC/FIS-AST/32113/2017 & POCI-01-0145-FEDER-032113; PTDC/FIS-AST/28953/2017 & POCI-01-0145-FEDER-028953. VA and SS acknowledge the support from FCT through Investigador FCT contracts of references IF/00650/2015 and IF/00028/2014, and POPH/FSE (EC) by FEDER funding through the program "Programa Operacional de Factores de Competitividade – COMPETE". This work of CJAPM was financed by FEDER—Fundo Europeu de Desenvolvimento Regional funds through the COMPETE 2020—Operational Programme for Competitiveness and Internationalisation (POCI), and by Portuguese funds through FCT—Fundação para a Ciência e a Tecnologia in the framework of the projects POCI-01-0145-FEDER-028987 and UID/FIS/04434/2019. OD is supported in the form of work contract (DL 57/2016/CP1364/CT0004). MRZO acknowledges financial support from AYA2016-79425-C3-2-P from the Spanish Ministry for Science, Innovation and Universities (MICIU). JIGH acknowledges financial support from the MICIU under the 2013 Ramón y Cajal program MICIU RYC-2013-14875. JIGH, RRL, CAP and ASM also acknowledge financial support from the MICIU for project AYA2017-86389-P. This publication makes use of The Data & Analysis Center for Exoplanets (DACE), which is a facility based at the University of Geneva (CH) dedicated to extrasolar planets data visualisation, exchange and analysis. DACE is a platform of the Swiss National Centre of Competence in Research (NCCR) PlanetS, federating the Swiss expertise in Exoplanet research. The DACE platform is available at https://dace.unige.ch. Based on observations made at ESO Very Large Telescope (Paranal, Chile) under programme 1102.C-744 and ESO 3.6 m telescope (La Silla, Chile) under programmes 090.C-0540 and 100.C-0750.

**AUTHOR CONTRIBUTIONS**

DE, CL and RA led the data analysis and interpretation. DE wrote the paper with contributions from RA. CL led the development of the data reduction pipeline. MZO coordinated the observations and scientific work and performed the first-epoch observation. FP, SC, RRL and NS led the ESPRESSO consortium and building of the instrument. JGH performed the second-epoch observation. FBor, OD, EPa, NS, EB, VB, HMC, NCB, JVS and HT brought decisive contributions to the interpretation. NCB performed an independent data analysis. SS performed the stellar parameter analysis. XD created the CCF mask and retrieved the list of its atomic lines. NH made the radial velocity retrieval. DS provided support with DACE. BL provided the nested sampling algorithm for the analysis. ML derived the transit ephemeris. VA, CA, YA, FBou, VD, PF, RGS, CM, AM, GM, PM, NN, GL, EPo, AS, ASM and SU participated to the scientific preparation and target selection for these observations. The other co-authors provided key contributions to the instrumental, software and operational development of ESPRESSO. All co-authors read and commented the manuscript.

## METHODS

### Observation log

The observations were carried out as part of the ESPRESSO Guaranteed Time Observation programme 1102.C-744. The observing conditions for both epochs (seeing, air masses, signal-to-noise ratios vs. time) are reported in Extended Data Fig. 1. The seeing at the beginning of the observations was better in epoch 2 (~0.9 arcsec) than in epoch 1 (~1.3 arcsec). This explains why we opted for a longer exposure time in epoch 1. We excluded exposures obtained at air mass above 2.2 as the atmospheric dispersion corrector (ADC) cannot handle correctly higher air masses.

### Stellar parameters

WASP-76 ($01^h46^m$, +02°42') is a F7 star of magnitude $V$=9.5. Its properties were studied in the discovery paper by ref. [17], who used photometry from SuperWASP-North, WASP-South, TRAPPIST, EulerCAM at the Euler telescope and spectroscopy from CORALIE at the Euler telescope and SOPHIE at the OHP 1.93m telescope. One spectroscopic transit was later observed with HARPS at the ESO 3.6m telescope on 11 Nov 2012[18]. Ref. [19] recently reported on two new HARPS spectroscopic transits (24 Oct 2017 and 22 Nov 2017), accompanied by simultaneous EulerCAM photometry. In the meantime, the star was observed by *Gaia*: ref. [19] used the *Gaia* DR2 values (parallax, magnitude and effective temperature) and the new EulerCAM photometry to re-assess the stellar parameters. This resulted in increased values for the stellar radius (hence, the radius of the planet) and the stellar mass compared to the ones previously reported[17,18], which are based on a combined analysis of the photometric *and* spectroscopic data.

We performed a new analysis of the stellar parameters based on the ESPRESSO spectra and the *Gaia* DR2 parallax ($\pi = 5.12 \pm 0.15$ mas). For this, we combined several of our spectra to obtain a high–signal-to-noise ratio spectrum of ~1,200 per resolution element, which we analysed using ARES+MOOG following a well-established spectroscopic analysis method[31-34]. The new stellar parameters we derive are listed in Extended Data Table 1. In particular, we obtain a stellar effective temperature of $6,329 \pm 65$ K compatible with ref. [18] ($6,250 \pm 100$ K), and a log $g$ of $4.196 \pm 0.106$ dex. Using the Padova stellar model isochrones (http://stev.oapd.inaf.it/cgi-bin/param_1.3) [35,36] and the *Gaia* parallax, we obtain a stellar age of $1.816 \pm 0.274$ Gyr, a *B-V* colour of $0.569 \pm 0.017$ mag, a stellar mass of $1.458 \pm 0.021\ M_\odot$ and a stellar radius of $1.756 \pm 0.071\ R_\odot$. The two latter values, strongly constrained by the Gaia

**Binary companion**

WASP-76 has a candidate companion separated by 0.4438±0.0053 arcsec and a magnitude difference of 2.58±0.27[37,38], which corresponds to a flux contrast of ~10. By combining these ground-based measurements with Keck and *HST*/STIS images, it is possible to establish that the candidate companion is actually bound to the star (G. Fu, private communication) and determine its effective temperature (~5,100 K) and radius ( 0.8 $R_\odot$ ). WASP-76B thus resembles a late G- or early K-type dwarf. This candidate companion lies at the limit of the entrance of the 0.5 arcsec-radius ESPRESSO fibre. Given the seeing values reported in Extended Data Fig. 1a,d, it most certainly contaminates our spectra and the contamination could vary with the seeing. We checked that the companion does not impact the stellar parameters by repeating our analysis selecting stellar spectra only in conditions of good seeing (≤ 0.85") or bad seeing (> 1"). The results are compatible with the values derived above within their stated uncertainties. We performed an extensive search for a spectroscopic signature of WASP-76B; however, we could not find any sign of a contamination of the stellar CCFs of WASP-76A down to the ~500 ppm level. Given its above-stated properties, it is surprising that the CCF of WASP-76B remain undetectable. One possibility is that WASP-76B is a fast rotator, producing a broad CCF that would be lost in the noise; however, the system is not particularly young. Another possibility is that the CCF of WASP-76B has almost identical radial velocity, FWHM and contrast as the CCF of the primary star, that would thus efficiently "hide" it. Because it is so well hidden, WASP-76B is unlikely to affect our results (see also "Transit photometry").

**Transit photometry**

We performed a new photometric analysis based on all six existing transit light curves of WASP-76b obtained with the EulerCam instrument at the Swiss Euler 1.2 m telescope in La Silla, Chile. We extracted the raw light curves (Extended Data Fig. 7a) using aperture photometry described in ref. [39]. We jointly analysed all photometric data sets using a differential-evolution Markov chain Monte Carlo code[40], fitting for the mid-transit time, revolution period, planet-to-star radius ratio and system scale with Gaussian priors. We accounted for instrumental systematics and red noise (see ref. [19] for more details). We used a quadratic limb-darkening law and obtained the corresponding $u_1$ and $u_2$ coefficients for WASP-76 with the routines of ref. [41]. The corrected light curves are shown in Extended Data Fig. 7b and derived parameters are reported in Extended Data Table 1. These values take into account the dilution caused by WASP-76B. Its main effect is to increase slightly the transit depth and planet-to-star radius ratio. However, the effect of the planet radius is small; we found $1.863^{+0.070}_{-0.083}$ $R_{2l}$. This value is actually smaller than the one previously reported by ref. [19]; this is because we made use of a smaller (and more accurate) stellar radius (see "Stellar parameters"). We used our newly derived parameters to create a transit model[42] that we utilised to perform the reloaded Rossiter-McLaughlin analysis (see "Reloaded Rossiter-McLaughlin effect and Doppler shadow") and extract to convert the residual amplitudes into differential transit depths (Extended Data Fig. 6c).

**Cross-correlation mask**

We used the built-in cross-correlation mask corresponding to an F9-type star in the ESPRESSO pipeline to obtain the stellar CCF for each exposure. This mask was created by collecting the position of all lines for the F9.5 star HD1581. The lines were individually identified by querying their wavelengths in the Vienna Atomic Line Database (VALD; http://vald.astro.uu.se). Iron (Fe) is by far the most represented element (47% of the 4,653 spectral lines). The following most-represented atoms are nickel (Ni, 6.5%), chromium (Cr, 5.7%) and titanium (Ti, 4.8%). The most represented ion in the F9 mask is $Ti^+$ (2.8% of the lines). The CCF is then built as a flux-weighted and contrast-weighted mean line profile. Many deep (contrasted) Fe lines are located in spectral regions where the stellar (continuum) flux is high, therefore boosting the importance of the Fe lines. Their actual weight in the F9 mask represents 76% of the spectroscopic information.

**Orbital solution**

We retrieved the orbital parameters and planet mass from our ESPRESSO measurements. We excluded data points obtained during transits from the analysis to prevent the Rossiter-McLaughlin effect perturbing the orbital solution. This left most of our in-transit data aside, hence we collected new data points to extend our coverage of the orbit (in particular, at the quadratures) at high precision. The new radial velocities are presented in Extended Data Table 2 and Extended Data Fig. 2.

To derive the uncertainties on the orbital parameters from the ESPRESSO data, we computed their posterior distribution with a Monte Markov Chain (MCMC) algorithm. We modelled the signal with a Keplerian and three radial velocity offsets: offsets 1 and 2 correspond to epochs 1 and 2, respectively. A technical intervention on ESPRESSO occurred between these two epochs and could have changed the reference "zero" radial velocity of the spectrograph; hence the need to introduce an offset. Offset 3 corresponds to the set of subsequent observations (spanning BJD 2,458,684 to 2,458,754) obtained to increase the precision of the radial velocity solution. Our variables are these three offsets, the period $P$, the semi-amplitude $K_\star$ the eccentricity $e$, the argument of periastron $\omega$ and the inferior conjunction time $T_{\text{conj}}$. We parametrised the noise in the covariance matrix by three terms: two standard deviations for white ($\sigma_W$) and red ($\sigma_R$) noise and a characteristic timescale $\tau$ of the red noise. The prior distributions on all parameters are listed in Extended Data Table 3. The constraints on the period $P$ and $e \cos \omega$ are obtained from the *Spitzer*[25] and Euler transits (see "Transit photometry"). The MCMC algorithm is an adaptive Metropolis algorithm[43] as implemented in ref. [44]. Efficient numerical methods[45] can perform the inversion of matrices built as in Equation (2). We used a homegrown package (`spleaf`; Delisle et al. submitted) based on such a method to speed up the covariance calculations. To check the convergence of the chain, we computed the number of effective samples from the autocorrelation function of the chain[44,46]. We obtained the posterior distribution of the planet mass by computing it as a function of $P$, $K_\star$, $e$, the inclination $i_\star$ and the stellar mass $M_\star$. We used the MCMC samples of $P$, $K_\star$, $e$ and sampled independently $i_\star$ and $M_\star$ from their constraints as given in Extended Data Table 1. The distribution of $i_\star$ is approximated by a mixture of two Gaussians (with mean $\mu = 89.623°$, and standard deviations $\sigma_1 = 0.034°$ and $\sigma_2 = 0.005°$). The distribution of $M_\star$ is approximated by a Gaussian distribution ($\mu = 1.458\ M_\odot$, $\sigma = 0.021\ M_\odot$). The maximum-likelihood fit, the posterior median and the 1-$\sigma$ confidence intervals are given in Extended Data Table 3. The corner plot, made with the `corner.py` code[47] is presented in Extended Data Figure 3 and the radial velocity data with the maximum-likelihood fit in Extended Data Figure 2.

The maximum-likelihood of the eccentricity and its posterior median are close to 0. Previous studies based on CORALIE data have adopted a null eccentricity[17,18] and our more precise ESPRESSO measurements also points towards a circular orbit. The best constraint comes from the *Spitzer* measurement of $e \cos \omega = -0.00135 \pm 0.00083$[25], which also gives a strong indication towards a null eccentricity for the most likely (small) values of $\omega$. A null eccentricity is also strongly favoured by theory, considering the expected short circularisation time scale: an equilibrium tide[48] would damp an eccentricity of 10% in ~30 Myr. Given the age of the star (1.8±0.3 Gyr), there were ample time for the orbit to fully circularise, especially considering that a dynamical tide in the fluid layers of the planet would result in a higher dissipation factor. Note that the stellar tide could potentially excite the eccentricity if its rotation is fast enough (if the spin of the star is larger than 18/11 of the orbital frequency; Refs. [49,50]). This is not the case here since the stellar rotation frequency (~0.03 d$^{-1}$; see "Spin-orbit angle and stellar rotation") is much smaller than $18/11 \times P^{-1} = 0.9$ d$^{-1}$. This was also probably the case in the past for much more time than 30 Myr (see Fig. 2 in ref. [51]). Considering all of this, we decided to fix the eccentricity to 0.

**Reloaded Rossiter-McLaughlin effect and Doppler shadow**

The idea of the reloaded Rossiter-McLaughlin effect is to directly track the stellar surface radial velocity behind the transiting planet[22,52,53]. Following ref. [22], we shifted the stellar CCFs into the stellar rest frame. For this, we made use of the orbital solution obtained above. Since ESPRESSO observations are not flux-calibrated, the continuum levels of the CCFs are arbitrary. We normalised each CCF by its continuum level determined from the Gaussian fit. We then scaled each normalised CCF according to its timing with respect to the planetary transit. For this, we calculated the theoretical transit light curve as described in the "Transit photometry" section. We shifted in velocity the rescaled CCFs by the velocity measured for a mean "master" out-of-transit CCF. This is done so that the final result is independent of the velocity offsets (systemic velocity as well as instrument offsets discussed in "Orbital solution"). Finally, we produced the CCFs of the occulted stellar surface ("local" CCFs) by subtracting each scaled in-transit CCF from the master out-of-transit CCF.

The projected velocities of the stellar surface behind the planet during the transit appears first blueshifted, then redshifted, which indicates a prograde planetary orbit. The surface velocities roughly follow a straight line, indicative of solid-body rotation. We verified this using a dedicated stellar rotation models described below, which we adjusted to the data, at the exception of the flattened portion seen after mid-transit, at a time when the planet shimmer intersects with the Doppler shadow. We also excluded local CCFs where the stellar line was not detected at more than 5σ; these CCFs have been obtained at the ingress and egress.

**Spin-orbit angle and stellar rotation**

We fitted the stellar surface velocities with a model of stellar surface rotation assuming solid-body rotation[22]. For WASP-76, there is a known degeneracy between the projected spin-orbit angle $\lambda$, the projected equatorial rotational velocity of the star $v_{eq} \sin i_\star$ (where $i_\star$ is the inclination of the stellar spin with respect to the plane of the sky) and the impact parameter[18]he very small value of the impact parameter (the transit is almost central). The impact parameter can be expressed as $a/R_\star \cos i_p$, where $i_p$ is the inclination of the planetary orbit. We chose $\lambda$, $v_{eq} \sin i_\star$, $a/R_\star$ and $i_p$ as free parameters. We embedded the model in a nested sampling retrieval algorithm[54] to efficiently explore the full parameter space. The priors on the four parameters were set as: (i) a uniform prior on $\lambda$ ranging from −180° to 180°, (ii) a Gaussian prior ($\mu = 1.61$ km s$^{-1}$, $\sigma = 0.28$ km s$^{-1}$) on $v_{eq} \sin i_\star$, which we derived as the quadratic difference between the FWHM of the stellar local master CCF and the FWHM of the stellar master-out CCF; (iii, iv) the posterior distributions of the Euler photometry (see "Transit photometry") were chosen as priors for $a/R_\star$ and $i_p$.. We performed a run of 5,000 living points; the best-fit parameters are the ones maximising the logarithm of the evidence, $\log \mathcal{Z}$. The maximum $\log \mathcal{Z}$ of 8.59±0.04 was obtained for $\lambda = 61.28^{+7.61}_{-5.06}$ deg, $v_{eq} \sin i_\star = 1.48 \pm 0.28$ km s$^{-1}$, $a/R_\star = 4.09 \pm 0.07$ and $i_p = 89.74^{+0.15}_{-0.11}$. The quoted 1σ uncertainties are obtained from the posterior distributions of the parameters, which are shown in Extended Data Fig. 4. Based on the value of $\lambda$ and the host star effective temperature of 6,329 K, WASP-76b lies at the transition between aligned and misaligned hot gas giants[55]. The slow (projected) rotation velocity we derived hints at a non-negligible inclination $i_\star$ of the stellar spin axis towards the line of sight (which is not constrained by the Rossiter-McLaughlin effect). In fact, $i_\star = 90°$, would yield a rotation period of 60 days, which is much larger than the typical range of ~15-40 days expected for F stars{2015MNRAS.452.2745S, SuarezMascareno:2016co}. A light curve of WASP-76 obtained with the All Sky Automated Survey for Supernovæ (ASAS) hints at a periodicity of ~35 days, which would yield an inclination of $i_\star \approx 36°$ with respect to the line of sight, i.e. the star would be close to pole-on.

**Constraints on the temperature rise across the evening terminator of an ultra-hot gas giant**
Ultra-hot gas giants are an emerging class of exoplanets. In addition to WASP-76b, some of its representative objects are WASP-12b[56], WASP-33b[1], WASP-103b[57], WASP-121b[58], MASCARA-2b[59] and KELT-9b[2]. Observations have enabled the measurement of some of their physical and chemical properties, such as their temperature structures or composition[3,10-14,19,60,61]; however, no consistent picture of these extreme climates exists yet, as interpretations are essentially based on global circulation models established for less-irradiated hot gas giants [7,8,15,26,27,62,63]. While these models can be used to interpret wind measurements in planets like HD 209458b[16] or HD 189733b[64], they are less adapted to objects like WASP-76b. Recent theoretical developments towards understanding ultra-hot atmospheres[4-6,9,65] and future works will be the basis to finely interpret spectroscopically- and temporarily-resolved measurements such as the ones presented here.

In particular, we exploit here the idea that atoms recombine into molecules across the evening terminator and molecules dissociate into atoms across the morning terminator of an ultra-hot gas giant[9]. Applying this to iron atoms make it possible to use the Fe signature as a thermometer. The lowest absorption depth of $\delta_{atm}^{eve} = \mathbf{434 \pm 32}$ ppm is measured when only the evening terminator contributes to the signal (between steps 1 and 2 of the scenario depicted in Fig. 3). This absorption depth can be expressed as $\delta_{atm}^{eve} \approx 2(R_p/R_\star)^2 (H^{eve}/R_p) n_H$, where $H^{eve}$ is the atmospheric scale height at the evening terminator and $n_H$ is the number of scale heights over which the absorption takes place. The largest absorption depth of $\delta_{atm}^{day} = \mathbf{628 \pm 49}$ ppm is observed at the end of the transit, when the line-of-sight through the trailing limb probes regions close to the dayside hot spot (step 3 in Fig. 3). Assuming the absorption signal takes place over the same number of scale heights as on the evening terminator, we can write that $\delta_{atm}^{day}/\delta_{atm}^{eve} = H^{day}/H^{eve}$. Since $H = kT/\mu g$, where $k$ is Boltzmann's constant, $\mu$ is the mean molecular mass of the atmosphere, $g$ is the surface gravity and $T$ is the temperature, it comes that $\delta_{atm}^{day}/\delta_{atm}^{eve} = T^{day}/T^{eve} \approx 1.5 \pm 0.2$. An occultation of WASP-76b by its host star was observed with *Spitzer*, providing a measurement of the dayside brightness temperature at 3.6 μm: 2,693±56 K[25]. Using this value and the variation of the absorption depth measured here, we can estimate that WASP-76b has an evening terminator temperature of 1,795±242 K.

**Competing interests**

The authors declare no competing interests.

**Data availability**

The data that support the findings of this study are available on request from the corresponding author (DE). The data are not publicly available due to the proprietary status of data obtained in the framework of the ESPRESSO Guaranteed Time Observations. At the end of the proprietary period, the data will be publicly available on the ESO archive (archive.eso.org).

**Code availability**

The ESPRESSO DRS is public software available from ESO at https://www.eso.org/sci/software/pipelines/espresso/espresso-pipe-recipes.html. The main analysis routines have been written by the authors in Interactive Data Language and are available upon reasonable request from the corresponding author (DE).

**Extended Data Table 1 | Parameters for WASP-76 and its planet.**

| Parameter | Unit | Value | Reference/Methods |
|---|---|---|---|
| *Gaia* DR2 ID | – | 2512326349403275520 | CDS Simbad |
| Right ascension (J2000) | hms | 01$^h$46$^m$31.9$^s$ | CDS Simbad |
| Declination (J2000) | dms | +02°42'02.0" | CDS Simbad |
| $V$ | mag | 9.52±0.03 | CDS Simbad |
| Spectral type | – | F7 | CDS Simbad |
| Systemic velocity, $\gamma_{sys}$ | km s$^{-1}$ | –1.11±0.50 | CDS Simbad |
| Parallax, $\pi$ | mas | 5.12±0.16 | *Gaia* DR2 |
| Stellar properties derived from ESPRESSO spectra | | | |
| Stellar mass, $M_\star$ | M$_\odot$ | 1.458±0.021 | "Stellar parameters" |
| Stellar radius, $R_\star$ | R$_\odot$ | 1.756±0.071 | "Stellar parameters" |
| Effective temperature, $T_{eff}$ | K | 6,329±65 | "Stellar parameters" |
| Stellar surface gravity, log $g$ | cgs | 4.196±0.106 | "Stellar parameters" |
| Turbulent velocity, $v_{turb}$ | km s$^{-1}$ | 1.543±0.027 | "Stellar parameters" |
| Colour, $B–V$ | mag | 0.569±0.017 | "Stellar parameters" |
| Metallicity, [Fe/H] | – | 0.366±0.053 | "Stellar parameters" |
| Age | Gyr | 1.816±0.274 | "Stellar parameters" |
| Projected equatorial rotational velocity, $v_{eq} \sin i_\star$ | km s$^{-1}$ | 1.48±0.28 | "Spin-orbit angle" |
| Spin-orbit projected angle, $\lambda$ | deg | $61.28^{+7.61}_{-5.06}$ | "Spin-orbit angle" |
| System properties retrieved from radial velocities | | | |
| Eccentricity, $e$ | – | 0 (fixed) | "Orbital solution" |
| Semi-amplitude of the stellar RVs, $K_\star$ | m s$^{-1}$ | $116.02^{+1.29}_{-1.35}$ | "Orbital solution" |
| Planet mass, $M_p$ | $M_{2\!\!\!|}$ | $0.894^{+0.014}_{-0.013}$ | "Orbital solution" |
| Systemic velocity for epoch 1, $\gamma_{181}$ | m s$^{-1}$ | $-1,162.00^{+2.86}_{-2.63}$ | "Orbital solution" |
| Systemic velocity for epoch 2, $\gamma_{182}$ | m s$^{-1}$ | $-1,167.54^{+2.79}_{-2.73}$ | "Orbital solution" |
| Systemic velocity for epoch 3, $\gamma_{193}$ | m s$^{-1}$ | $-1,171.11^{+1.28}_{-1.36}$ | "Orbital solution" |
| System properties retrieved from photometry | | | |
| Period $P$ | days | $1.80988198^{+0.00000064}_{-0.00000056}$ | "Transit photometry" |
| Mid-transit time $T_c$ | BJD | $58080.626165^{+0.000418}_{-0.000367}$ | "Transit photometry" |
| Radius ratio $R_p/R_\star$ | – | $0.10852^{+0.00096}_{-0.00072}$ | "Transit photometry" |
| System scale $a/R_\star$ | – | $4.08^{+0.02}_{-0.06}$ | "Transit photometry" |
| Inclination $i$ | deg | $89.623^{+0.005}_{-0.034}$ | "Transit photometry" |
| Phases of contacts I and IV, $\varphi_{1-4}$ | – | ±0.043 | "Transit photometry" |
| Phases of contacts II and III, $\varphi_{2-3}$ | – | ±0.034 | "Transit photometry" |
| Ingress duration, $\Delta T_{12}$ | min | 23.6 | "Transit photometry" |
| Total transit duration, $\Delta T_{14}$ | min | 230 | "Transit photometry" |
| Transit depth | % | $1.178^{+0.077}_{-0.076}$ | "Transit photometry" |
| Semi-major axis, $a$ | au | 0.0330±0.0002 | "Transit photometry" |
| Impact parameter, $b$ | – | $0.027^{+0.13}_{-0.023}$ | "Transit photometry" |
| Quadratic limb-darkening coefficient $u_1$ | – | 0.393 | "Transit photometry" |
| Quadratic limb-darkening coefficient $u_2$ | – | 0.219 | "Transit photometry" |
| Combined parameters | | | |
| Semi-amplitude of the planet RVs, $K_p$ | km s$^{-1}$ | 196.52±0.94 | This work |
| Planet radius, $R_p$ | $R_{2\!\!\!|}$ | $1.854^{+0.077}_{-0.076}$ | "Transit photometry" |
| Planet density, $\rho_p$ | g cm$^{-3}$ | 0.17±0.02 | This work |
| Planet surface gravity, $g_p$ | m s$^{-2}$ | 6.4±0.5 | This work |
| Total stellar irradiance | $\mathcal{S}^N_\odot$ | 4,104±896 | This work |
| Equilibrium temperature for null albedo | K | 2,228±122 | This work |
| Dayside brightness temperature at 3.6 μm | K | 2,693±56 | Ref. [25] |
| Atmospheric scale height (dayside) | km | 1,501±130 | This work |
| Differential transit depth of one scale height | ppm | 266±26 | This work |

**Extended Data Table 2 | Radial velocities of WASP-76 obtained with ESPRESSO.**

Offsets 1, 2 and 3 have been applied to epochs 1 (2018-09-02), 2 (2018-10-30) and 3 (Fall 2019), respectively. These data exclude the points obtained during transit.

| BJD | RV (m s$^{-1}$) | $\sigma_{RV}$ (m s$^{-1}$) | Epoch |
|---|---|---|---|
| 58364.65995 | -1114.88 | 1.67 | 1 |
| 58364.66760 | -1116.34 | 1.39 | 1 |
| 58364.67561 | -1123.08 | 1.00 | 1 |
| 58364.68299 | -1124.66 | 0.79 | 1 |
| 58364.69069 | -1131.50 | 0.85 | 1 |
| 58364.86077 | -1194.98 | 0.55 | 1 |
| 58364.86842 | -1198.83 | 0.56 | 1 |
| 58364.87606 | -1200.42 | 0.58 | 1 |
| 58364.88374 | -1200.96 | 0.72 | 1 |
| 58364.89110 | -1201.34 | 1.00 | 1 |
| 58364.89953 | -1206.00 | 0.97 | 1 |
| 58364.90687 | -1209.20 | 0.91 | 1 |
| 58364.91482 | -1212.37 | 0.86 | 1 |
| 58364.92248 | -1215.44 | 0.83 | 1 |
| 58422.55814 | -1114.88 | 1.04 | 2 |
| 58422.56232 | -1115.76 | 1.01 | 2 |
| 58422.56641 | -1118.47 | 0.99 | 2 |
| 58422.57082 | -1119.13 | 1.01 | 2 |
| 58422.57510 | -1122.64 | 0.95 | 2 |
| 58422.57924 | -1124.56 | 0.90 | 2 |
| 58422.58352 | -1126.14 | 0.91 | 2 |
| 58422.58780 | -1126.88 | 0.90 | 2 |
| 58422.59203 | -1126.98 | 0.90 | 2 |
| 58422.59621 | -1130.71 | 0.94 | 2 |
| 58422.60051 | -1131.66 | 1.02 | 2 |
| 58422.60477 | -1134.06 | 0.95 | 2 |
| 58422.60907 | -1135.32 | 0.93 | 2 |
| 58422.61325 | -1136.36 | 0.86 | 2 |
| 58422.77451 | -1201.85 | 0.88 | 2 |
| 58422.77879 | -1201.62 | 0.85 | 2 |
| 58422.78295 | -1201.66 | 0.86 | 2 |
| 58422.78722 | -1208.19 | 0.92 | 2 |
| 58422.79147 | -1206.84 | 0.85 | 2 |
| 58422.79571 | -1209.68 | 0.84 | 2 |
| 58422.79999 | -1211.27 | 0.83 | 2 |
| 58422.80421 | -1210.52 | 0.85 | 2 |
| 58422.80857 | -1216.66 | 0.89 | 2 |
| 58422.81266 | -1213.72 | 0.89 | 2 |
| 58422.81693 | -1218.75 | 0.94 | 2 |
| 58422.82116 | -1216.22 | 0.96 | 2 |
| 58422.82543 | -1222.85 | 0.95 | 2 |
| 58422.82970 | -1218.60 | 0.98 | 2 |
| 58422.83398 | -1223.08 | 0.90 | 2 |
| 58422.83819 | -1222.37 | 0.90 | 2 |
| 58422.84241 | -1222.71 | 0.94 | 2 |
| 58422.84661 | -1224.40 | 1.01 | 2 |
| 58684.91813 | -1095.81 | 0.69 | 3 |
| 58684.92351 | -1096.32 | 0.71 | 3 |
| 58695.85571 | -1119.54 | 1.01 | 3 |
| 58696.92344 | -1152.16 | 1.50 | 3 |
| 58706.93572 | -1208.77 | 0.77 | 3 |
| 58719.85812 | -1283.64 | 0.63 | 3 |
| 58721.85408 | -1285.71 | 0.72 | 3 |
| 58725.88068 | -1157.62 | 1.02 | 3 |
| 58731.80455 | -1061.76 | 0.69 | 3 |
| 58741.64369 | -1284.85 | 0.64 | 3 |
| 58741.65148 | -1286.11 | 0.60 | 3 |
| 58741.65905 | -1286.98 | 0.58 | 3 |
| 58741.66666 | -1286.84 | 0.56 | 3 |
| 58752.63975 | -1282.47 | 5.69 | 3 |
| 58752.64642 | -1279.18 | 1.81 | 3 |
| 58752.65049 | -1276.87 | 1.79 | 3 |
| 58752.65472 | -1275.84 | 1.77 | 3 |
| 58752.65877 | -1279.07 | 1.94 | 3 |
| 58752.66323 | -1275.61 | 1.84 | 3 |
| 58752.66730 | -1273.82 | 1.75 | 3 |
| 58752.67154 | -1275.40 | 1.87 | 3 |
| 58752.67574 | -1270.26 | 2.06 | 3 |
| 58752.67983 | -1273.41 | 1.83 | 3 |
| 58752.68687 | -1272.35 | 1.19 | 3 |
| 58752.69488 | -1268.70 | 0.99 | 3 |
| 58752.70012 | -1253.64 | 21.49 | 3 |
| 58753.67498 | -1086.73 | 0.95 | 3 |
| 58754.79119 | -1177.80 | 0.70 | 3 |

**Extended Data Table 3 | Orbital elements from the MCMC retrieval on the radial velocities.**

| Parameter | Unit | Prior | Maximum likelihood | Posterior median |
|---|---|---|---|---|
| Period $P$ | days | Gaussian ($\mu$=1.80988198 d, $\sigma$=6.4×10$^{-7}$ d) | 1.8098821 | 1.8098819(7) |
| Semi-amplitude $K_\star$ | m s$^{-1}$ | Uniform on [0,30] km s$^{-1}$ | 115.94 | $116.02^{+1.29}_{-1.35}$ |
| $T_{\mathrm{conj}} - T_c$ | s | Gaussian on $T_{\mathrm{conj}}$ ($\mu$=58,080.626165 BJD, $\sigma$=4.1×10$^{-4}$ d) | 37 | $5^{+24}_{-34}$ |
| $\sqrt{e}\cos\omega$ | – | Gaussian on e cos ω ($\mu$=−0.0013, $\sigma$=8×10$^{-4}$) | −0.0562 | $-0.0169^{+0.0132}_{-0.0102}$ |
| $\sqrt{e}\sin\omega$ | – | Uniform | 0.001 | $-0.062^{+0.092}_{-0.078}$ |
| $\sigma_W^2$ | m s$^{-1}$ | Truncated Gaussian on $\sigma_W^2$ ($\sigma$=100 m$^2$ s$^{-2}$) | 1.67 | $1.29^{+0.25}_{-0.28}$ |
| $\sigma_R^2$ | m s$^{-1}$ | Truncated Gaussian on $\sigma_R^2$ ($\sigma$=100 m$^2$ s$^{-2}$) | 1.41 | $3.13^{+1.01}_{-1.33}$ |
| Correlation time scale $\tau$ | days | Log-uniform on 1/$\tau$ on [0.001,1000] d | 0.04 | $1.090^{+2.56}_{-1.08}$ |
| Offset 1 | m s$^{-1}$ | Uniform on [−200,200] km s$^{-1}$ | −1,160.70 | $-1{,}162.00^{+2.86}_{-2.63}$ |
| Offset 2 | m s$^{-1}$ | Gaussian ($\mu$=offset 1, $\sigma$=10 m s$^{-1}$) | −1,167.78 | $-1{,}167.54^{+2.79}_{-2.73}$ |
| Offset 3 | m s$^{-1}$ | Gaussian ($\mu$=offset 1, $\sigma$=10 m s$^{-1}$) | −1,171.36 | $-1{,}171.11^{+1.28}_{-1.36}$ |
| Planet mass $M_p$ | $M_{♃}$ | | 0.894 | $0.894^{+0.014}_{-0.013}$ |

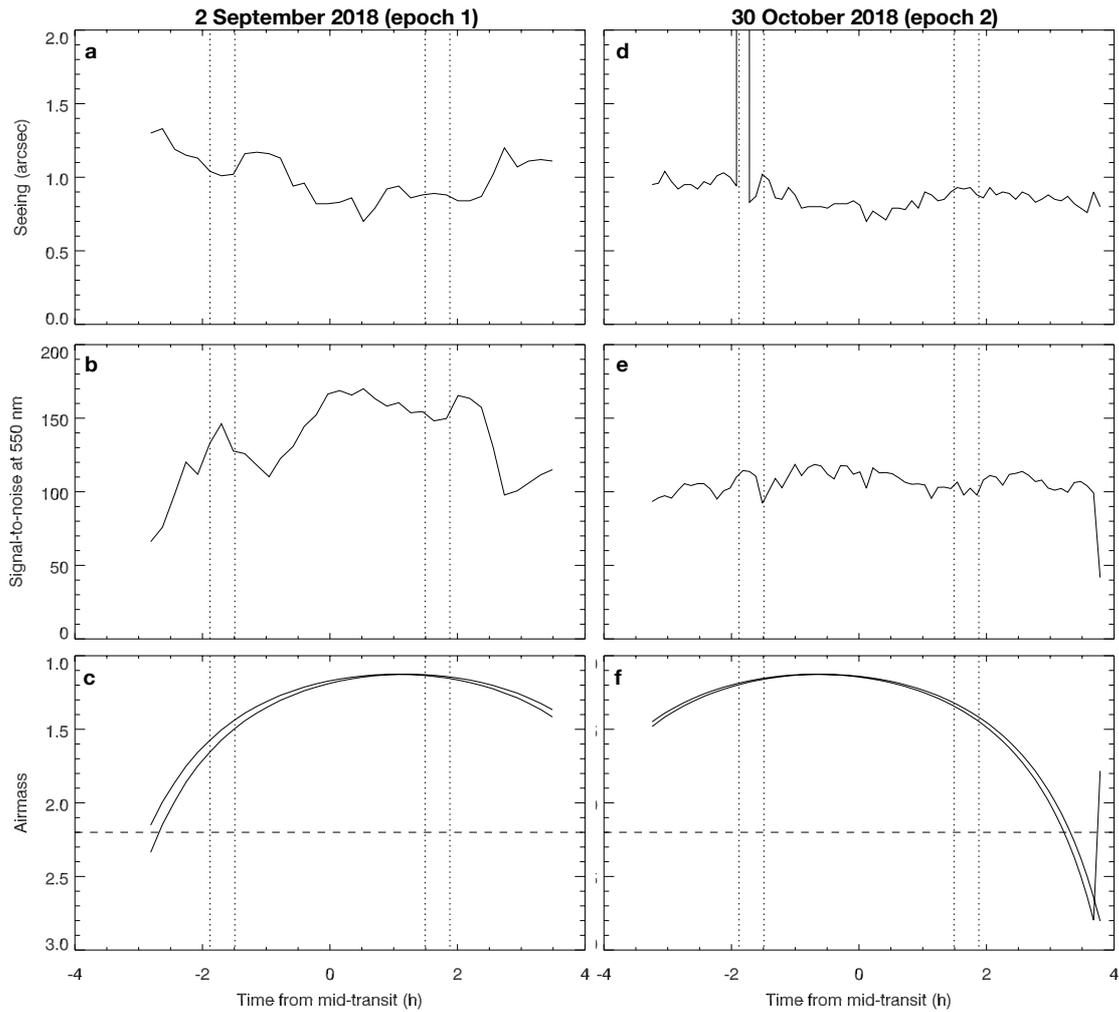

**Extended Data Figure 1 | Variations of the observing conditions** during transit epoch 1 (a,b,c) and epoch 2 (d,e,f). The seeing (a,d), signal-to-noise ratio per pixel at 550 nm (b,e) and airmass (c,f) are shown as a function of the time in transit. Vertical dotted lines represent the transit contacts. The horizontal dashed lines in panels c and f indicate the airmass of 2.2 beyond which the data are discarded from the analysis.

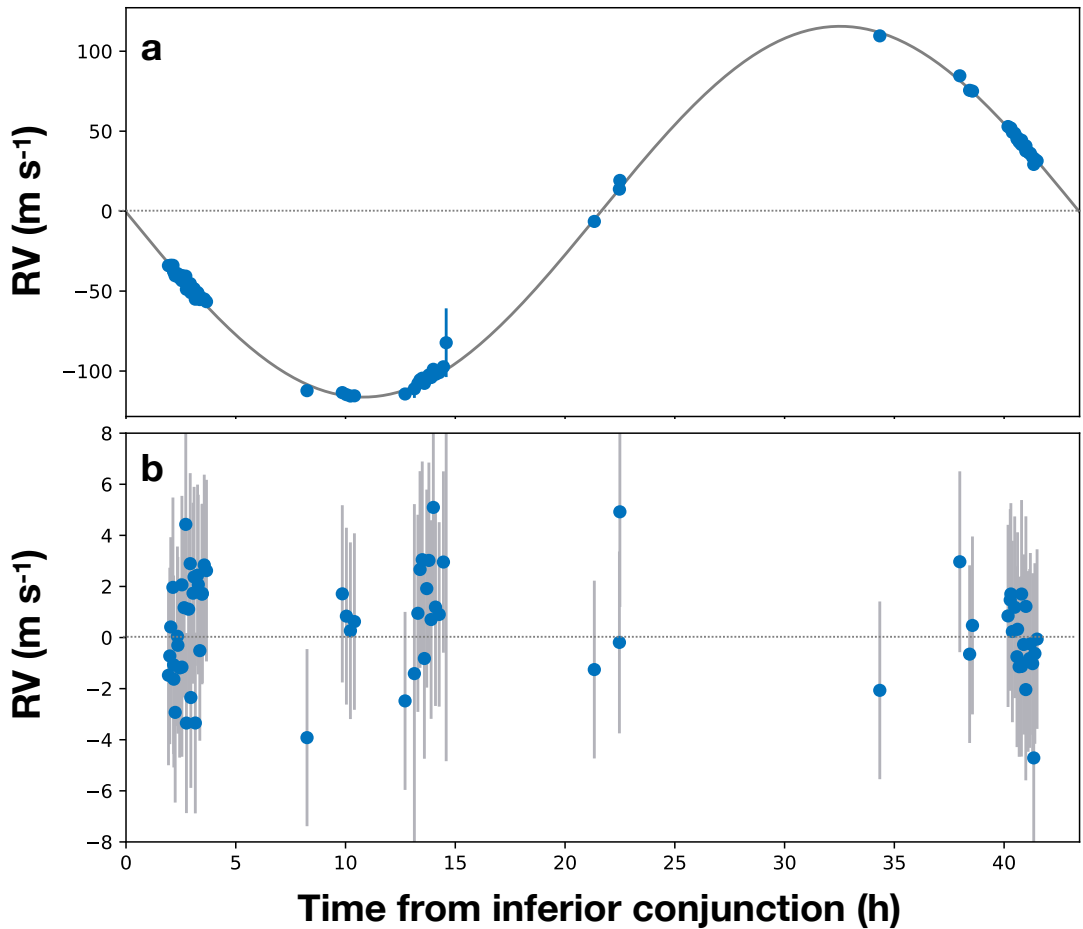

**Extended Data Figure 2 | ESPRESSO radial velocities of WASP-76. a**, Stellar radial velocities (blue points) and the maximum-likelihood fit using values from Extended Data Table 3. The transit occurs at the inferior conjunction (0 h). In-transit data have been removed as they are affected by the Rossiter-McLaughlin effect and the atmospheric absorption from the planet. **b**, Residuals of the radial velocities after subtraction of the maximum-likelihood fit. The standard deviation of the residuals is ~2.8 m s$^{-1}$.

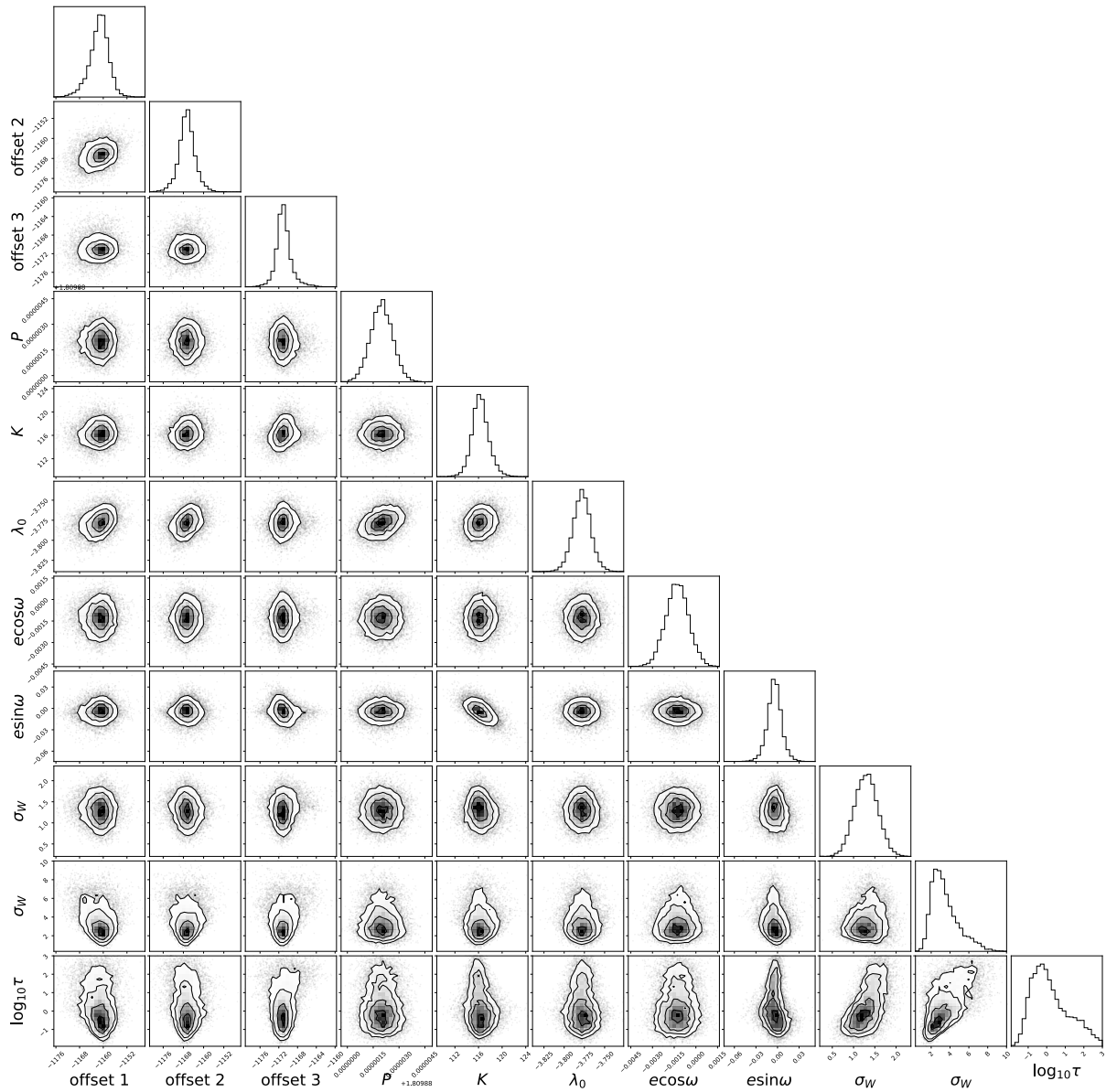

**Extended Data Figure 3 | MCMC chain corner plot for the orbital parameters** representing the posterior distribution of variables used for the MCMC computations of the orbital parameters. The posterior distribution medians are reported in Extended Data Table 3.

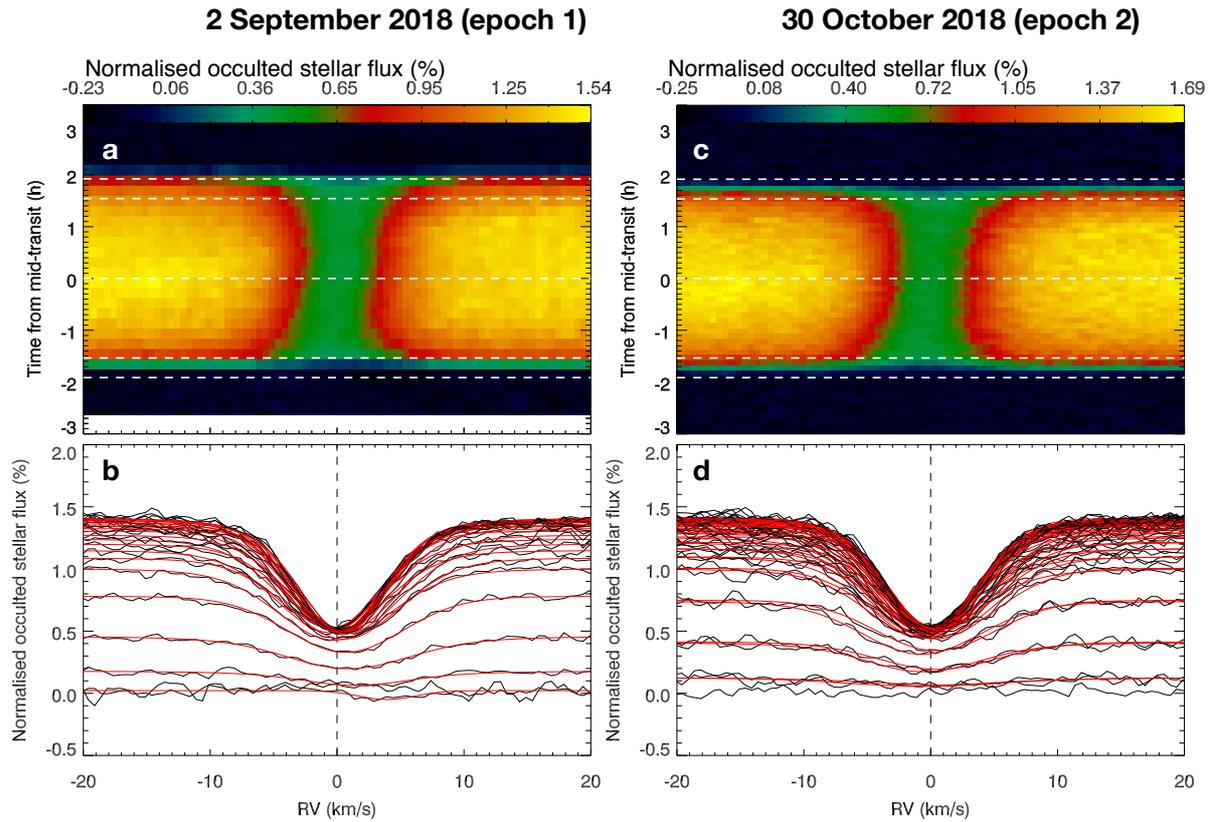

**Extended Data Figure 4 | Doppler shadow of WASP-76. a,c,** Local stellar CCFs behind the planet represented as a function of time for epoch 1 (a) and epoch 2 (c). The horizontal dashed lines represent (from bottom to top) the 2nd contact, mid-transit and 3rd contact. **b,d,** 1D view of the local stellar CCFs (black lines) with their Gaussian fits (red curves).

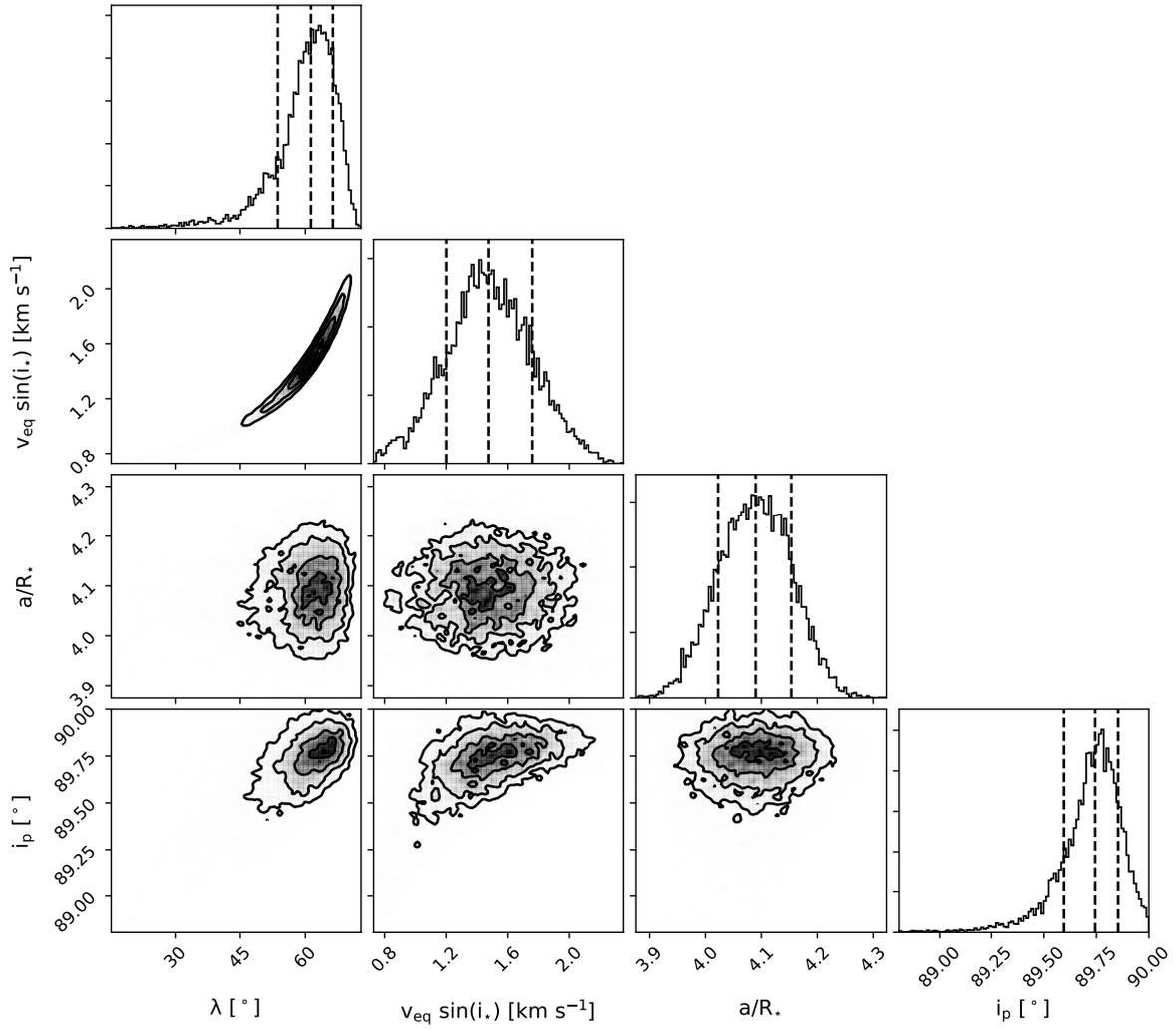

**Extended Data Figure 5 | Parameters of the stellar surface rotation model.** The corner plot shows the posterior distributions of the four free parameters of the model, the projected spin-orbit angle, $\lambda$, the projected equatorial stellar rotational velocity $v_{eq} \sin i_\star$, the system scale $a/R_\star$ and the planetary orbit inclination $i_p$. The posterior distribution medians and their 1σ uncertainties are represented by vertical dashed lines.

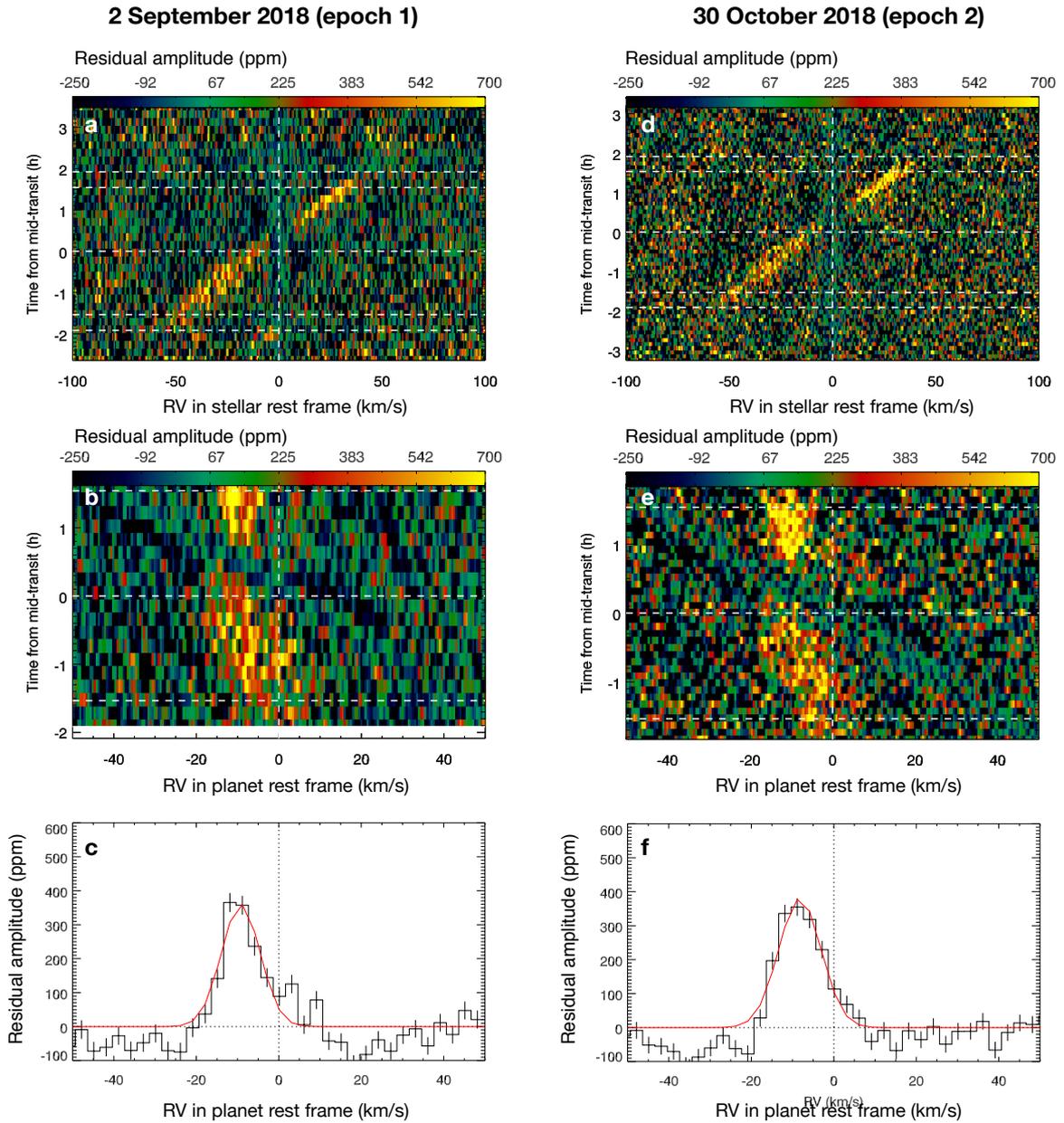

**Extended Data Figure 6 | Absorption signature of WASP-76b. a,b,c,** On 2 September 2018 (epoch 1). **d,e,f,** On 30 October 2018 (epoch 2). The planetary absorption signal is shown in the stellar rest frame (a,d), the planet rest frame (b,e) and is time-averaged in the planet rest frame to produce the atmospheric absorption profile integrated over the whole limb (c,f). An indicative Gaussian fit (red curves) is overplotted on the absorption profiles. Both epochs show compatible results.

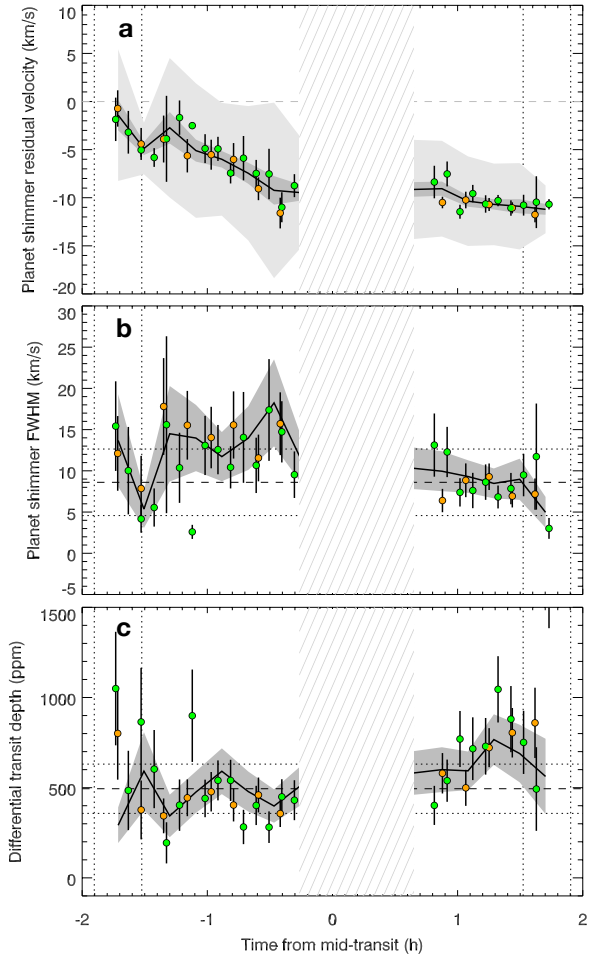

**Extended Data Figure 7 | Measured properties of the planetary absorption signature as a function of time.** Data from epoch 1 (orange), epoch 2 (green) and both epochs combined (binned by 2; black curve with 1σ uncertainty in dark grey) are shown. They result from Gaussian fits to the planetary absorption signal in the residual maps of Fig. 2b and Extended Data Figs. 5b and e. A factor of $(R_p/R_\star)^2/(1 - \Delta F/F(t))$ was applied to the residual maps before the fit, where $\Delta F/F(t)$ is the model light curve used to extract the Doppler shadow. **a,** Radial velocity of the planetary signal in the planet rest frame. The light grey region shows the FWHM associated to each point. **b,** The FWHM of the signal. The weighted-mean (horizontal dashed line) is 8.6±0.7 km s$^{-1}$. Horizontal dotted lines indicate the standard deviation of the values. **c,** Amplitude of the shimmer representing the differential transit depth. The weighted-mean is 494±27 ppm. The hatched area in all panels represents the overlap between the Doppler shadow and the planetary signal; data between −0.2 h and +0.7 h from mid-transit are excluded from the analysis.

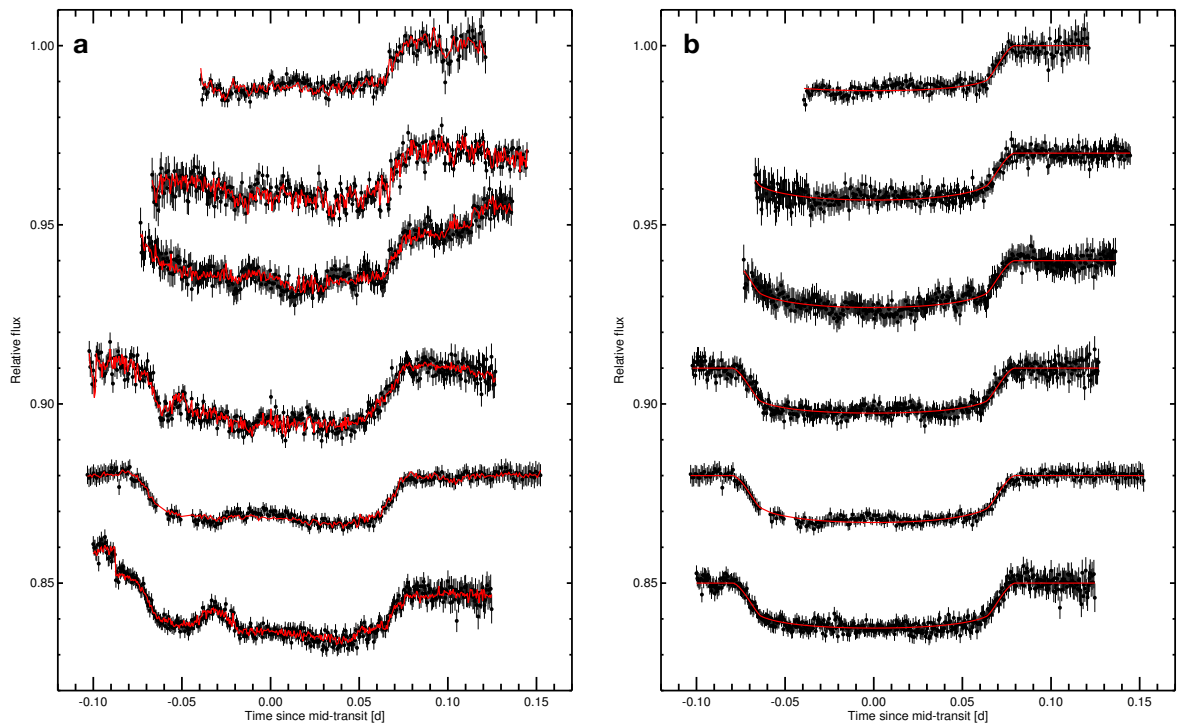

**Extended Data Figure 8 | Photometric transit light curve of WASP-76b** obtained with the EulerCam instrument on the Swiss Euler 1.2 m telescope in La Silla, Chile. The last three transits (bottom rows) have been previously reported in ref. [19]. **a**, Raw light curves with their best-fit models including systematic effects. **b**, Normalised light curves.